\documentclass[11pt,a4paper]{article}

\usepackage{jcappub}
\usepackage{graphicx}
\usepackage{color}
\usepackage{xspace}
\usepackage{relsize}
\usepackage{amssymb}
\usepackage{aas_macros}
\usepackage{fontenc}
\usepackage{hyperref}
\usepackage{amsmath}
\pdfoutput = 1

\newcommand*{\SONG}{\textsf{SONG}\xspace}

\newcommand*{\F} [2] {\ensuremath{F^{{#1},{#2}}}\xspace}

\newcommand*{\sref} [1] {Sec.\ \ref{#1}\xspace}

\newcommand*{\eref} [1] {Eq.~(\ref{#1})\xspace}

\newcommand*{\fref} [1] {Figure \ref{#1}\xspace}

\renewcommand*{\F} [2] {\ensuremath{F^{\,{#1},{#2}}}\xspace} 

\def\eprinttmp@#1arXiv:#2 [#3]#4@{{\href{http://arxiv.org/abs/#1}{#1}}}
\providecommand{\eprint}[1]{\eprinttmp@#1arXiv: [y]@}

\begin{document}

\title{A new line-of-sight approach to the non-linear Cosmic Microwave Background}


\author{Christian Fidler}
\affiliation{Institute of Cosmology and Gravitation, University of
Portsmouth, Portsmouth PO1 3FX, UK}
\author{Kazuya Koyama}
\affiliation{Institute of Cosmology and Gravitation, University of
Portsmouth, Portsmouth PO1 3FX, UK}
\author{Guido W. Pettinari}
\affiliation{Department of Physics \& Astronomy, University of Sussex,
Brighton BN1 9QH, UK}

\date{\today}

\abstract{
  
We develop the \emph{transport operator formalism}, a new line-of-sight integration framework to calculate the anisotropies of the Cosmic Microwave Background (CMB) at the linear and non-linear level.
This formalism utilises a transformation operator that removes all inhomogeneous propagation effects acting on the photon distribution function, thus achieving a split between perturbative collisional effects at recombination and non-perturbative line-of-sight effects at later times.
The former can be computed in the framework of standard cosmological perturbation theory with a second-order Boltzmann code such as \SONG, while the latter can be treated within a separate perturbative scheme allowing the use of non-linear Newtonian potentials.
We thus provide a consistent framework to compute all physical effects contained in the Boltzmann equation and to combine the standard remapping approach with Boltzmann codes at any order in perturbation theory, without assuming that all sources are localised at recombination.

}


\maketitle

\section{Introduction}
\label{introduction}
Recently, there has been growing interest in computing the second-order Cosmic Microwave Background (CMB) anisotropies
\cite{bartolo:2004a, bartolo:2004b, bartolo:2012a, senatore:2009a, khatri:2009a, nitta:2009a, creminelli:2004a, boubekeur:2009a,  lewis:2012a, creminelli:2011a}. The Einstein-Boltzmann equations at second order have been studied in great detail \cite{bartolo:2006a, bartolo:2007a, pitrou:2007a,pitrou:2009b, beneke:2010a,naruko:2013a} and numerical codes have been developed to predict non-Gaussianity of the CMB anisotropies \cite{pitrou:2010a, huang:2013a, pettinari:2013a, su:2012a, fidler:2014a, pettinari:2014b}.  Even without any primordial non-Gaussianity of the curvature perturbation, the CMB anisotropies are inevitably non-Gaussian due to non-linear physics at recombination and the non-linear nature of Einstein's gravity. The intrinsic bispectrum induced by non-linear dynamics contains interesting information about physics at recombination and subsequent non-linear propagation of photons through inhomogeneous space-time. The dominant contribution in the intrinsic bispectrum due to the correlation between lensing and the Integrated Sachs-Wolfe (ISW) effect \cite{spergel:1999a, goldberg:1999a, seljak:1999a} has been detected by the Planck satellite \cite{planck-collaboration:2013b}.
In the future, a high-resolution experiment including information from polarisation might be able to observe the intrinsic bispectrum induced by non-linear physics \cite{pettinari:2014b}.
Precise estimations of the intrinsic bispectrum are also important to extract information on the primordial non-Gaussianity in future experiments.

Using the second-order Boltzmann code \SONG \cite{pettinari:2013a,pettinari:2014a}, the intrinsic bispectra for temperature and polarisation were calculated including all physical effects except the late-time non-linear evolution after recombination \cite{pettinari:2014b} (see also \cite{huang:2013a, huang:2014a, su:2012a}). There are several difficulties in including the late-time non-linear effects. At second order, the Boltzmann equation contains terms written as a product of the first-order metric perturbations and the photon density. These terms describe the propagation effects of photons through inhomogeneous space-time, i.e. gravitational redshifts, time delay and lensing. If we treat these terms as a source in the line-of-sight integration \cite{seljak:1996a}, they are not multiplied by the visibility function and contribute to the integral until late times after recombination. Then the source terms contain an infinite number of multipole moments and it is computationally impractical to evaluate the line-of-sight integration. Another issue is that, at late times, density perturbations become non-linear, making the second-order perturbation theory inapplicable on small scales, even though the metric perturbations remain small. Thus the second-order treatment becomes invalid at late times.
The dominant late-time effect comes from lensing. In order to include lensing, a remapping approach is often used to express the observed temperature as a product of the temperature anisotropies at the last scattering surface and the deflection angle due to lensing at late times \cite{hu:2000b, lewis:2006a}. 

There have been several attempts to solve the above problems in the Boltzmann equation and connect the standard remapping approach with the second-order calculation of the Boltzmann equation. 
Ref.~\cite{huang:2014a} showed that it is possible to avoid the problem of solving the full hierarchy of the photon density by using a transformation of variables and integrations by parts in the line-of-sight integration. Ref.~\cite{su:2014a} showed how the remapping approach can be reproduced from the Boltzmann equation. 

In this paper we explore an alternative strategy of computing the various late-time line-of-sight propagation effects for photons. 
We start from the Boltzmann equation and derive a transformation removing the propagation terms from the Boltzmann hierarchy. Solving the equations for the transformation operator $\mathcal{J}$ turns out to be more efficient than treating lensing and time-delay in the Boltzmann hierarchy as the perturbation theory can be truncated at a lower order. In addition this method separates early-time effects from late-time ones and provides a consistent framework for numerical Boltzmann codes that are used to solve the equations until the time of recombination. At later times when perturbation theory breaks down non-linear Newtonian treatments are needed. This formalism glues both approximations together and clarifies the connections between lensing in the remapping approach and non-linear Boltzmann codes. Compared with earlier attempts \cite{huang:2014a,su:2014a}, the transport operator formalism is the most general, relaxing the assumption that all sources are localised at recombination and valid to any order in perturbation theory. While preparing our paper, we became aware of another approach to this problem using the line-of-sight integration along a full geodesic \cite{saito:2014a}. We will make a brief comment on the relation between the two approaches. 

This paper is structured as follows. In section~\ref{Formalism}, we present our formalism to define the transformation operator $\mathcal{J}$ that encodes the late-time propagation effects of photons in inhomogeneous space-time without the collision term. The formalism is extended to include the collision term in section~\ref{Collisions}. We discuss our approach to set initial conditions for this transformation operator to make the separation between physics at the recombination and late-time gravitational effects efficient. In section~\ref{Applications}, we apply our formalism and show how we recover the standard remapping approach from the Boltzmann equation and explain how we include the finite width of recombination. An extension to polarisation is briefly discussed in section~\ref{Polarisation}. Section~\ref{Conclusions} is devoted to conclusions. In Appendix A, we compute corrections to the angular power spectrum due to the redshift-lensing correlation using our formalism.

\section{Formalism} \label{Formalism}
Before dealing with the propagation terms in the full Boltzmann equation we analyse the collision-free transport of a distribution function that is given at some initial time. We specify an initial distribution function of photons $f(x,p)$, with the comoving momentum $p$. This picture could for example resemble the cosmic microwave background after recombination. 
The equations of motion for the distribution function are given by the Boltzmann equation, while the gravitational potentials are determined by the Einstein equation. 
The Boltzmann equation reads:
\begin{equation}
  \label{eq:boltzmann}
\frac{\partial}{\partial\eta} f + n^i \frac{\partial f}{\partial x^i} + \frac{dp}{d\eta}\frac{\partial f}{\partial p} + \frac{dn^i}{d\eta}\frac{\partial f}{\partial n^i} +  \Big(\frac{dx^i}{d\eta}-n^i \Big)\frac{\partial f}{\partial x^i} = 0  \;,
\end{equation}
where $\eta$ is the conformal time and $n^i$ is the direction of the comoving momentum $p$.
The first two terms describe the free propagation of photons while the third term describes the redshifting of photons. Lensing, which is the change of the direction of photons, is given by the fourth term while the last one encodes the time-delay effects \cite{hu:2000a}. We define the free propagation operator 
\begin{equation}
{\cal D} = \frac{\partial}{\partial\eta} + n^i \frac{\partial }{\partial x^i} \;,
\end{equation}
and the gravitational deflection operator 
\begin{equation}
  \label{eq:tau_definition}
\tau = \frac{dp}{d\eta}\frac{\partial }{\partial p} + \frac{dn^i}{d\eta}\frac{\partial }{\partial n^i} +  \Big(\frac{dx^i}{d\eta}-n^i \Big)\frac{\partial }{\partial x^i}\;.
\end{equation}
Using these operators the transport equation is given by:
\begin{equation}
\label{eq:eom}
{\cal D} f+ \tau f = 0\;.
\end{equation}

We will now derive a transformation that removes all the inhomogeneous transport terms from the Boltzmann equation. This transformation is described by an operator $\mathcal{J}$ acting on the distribution function: 
\begin{equation}
f = \mathcal{J} \tilde{f}\;.
\end{equation} 
Using the equations of motion \ref{eq:eom} we find for the transformed distribution function $\tilde{f}$: 
\begin{equation}
\mathcal{J}{\cal D}\tilde{f} + [{\cal D},\mathcal{J}]\tilde{f} + \tau \mathcal{J} \tilde{f} = 0\;,
\end{equation}
with the commutator of operators denoted by $[\cal{A},\cal{B}]=\cal{A} \cal{B}-\cal{B} \cal{A}$.
If the transformation operator $\mathcal{J}$ satisfies
\begin{equation}
[{\cal D},\mathcal{J}] + \tau \mathcal{J} = 0\;,
\end{equation}
the equations of motion for $\tilde{f}$ can be simplified to 
\begin{equation}
{\cal D} \tilde{f} = 0 \;,
\end{equation}
and all inhomogeneous propagation effects have been removed. The transformed distribution function evolves as in a unperturbed Universe, while all the complications of the non-linear transport are encoded in $\mathcal{J}$.  
Whether this is viable in practise depends on the complexity to find an operator $\mathcal{J}$ that satisfies the equation of motion $[{\cal D},\mathcal{J}] + \tau \mathcal{J} = 0$. Note that the operator $\mathcal{J}$ is independent of the distribution function and depends only on the structure of space-time. As this operator contains all information about the non-linear transport in a perturbed universe we will also refer to it as the transport operator.

The commutator $[{\cal D},\mathcal{J}]$ can be interpreted as a generalisation of the free propagation derivative for operators. 
Since $\mathcal{J}$ is sourced by $\tau$ which contains derivatives with respect to $n^i$ this commutator will be nontrivial. Instead of calculating the extra terms generated by $\mathcal{J}$ acting on $\cal{D}$, we expand $\mathcal{J}$ in a basis of operators which do commute with ${\cal D}$ and obtain equations for the coefficients in that basis. We rewrite 
\begin{equation}
  \label{eq:tau_parametrisation}
\tau \;=\; \tau_p \frac{\partial}{\partial p } + \tau_n^i \frac{\partial}{\partial n^i} + \tau_x^i \frac{\partial}{\partial x^i } \;=\; \tau_p \frac{\partial}{\partial p } + \tau_n^i \bar{D}_i + (\tau_x^i-\eta\tau_n^i) \frac{\partial}{\partial x^i } \;,
\end{equation}
with $\bar{D}_i = \frac{\partial}{\partial n^i} + \eta \frac{\partial}{\partial x^i}$. The angular derivative operator $\bar{D}_i$ is chosen such that $[\bar{D}_i, {\cal D}]= 0$; all operators of the basis $\partial_a = \{\frac{\partial}{\partial p },\frac{\partial}{\partial x^i },\bar{D}_i\}$ commute with ${\cal D}$, i.e. $[{\cal D},\partial_a]=0$. Expanding in this basis of differential operators ($\tau = \tau_a \partial_a$) we find the simple relation $[{\cal D},\tau] = {\cal D}(\tau_{a})\partial_{a}$. As $\mathcal{J}$ is sourced only by $\tau$, the commutator takes an analogous form for $\mathcal{J}$ discussed in \eref{eq:DofJ}. It is important to note that the price for simplifying the commutator is having a more complicated time-dependent operator basis.

While $\tau$ is linear in they differential basis, $\mathcal{J}$ may contain derivatives of higher order, complicating the expansion. We expand $\mathcal{J}$ using the basis  $\partial_a$
\begin{equation}
\mathcal{J}_{a...c}\,\partial_{a...c} \;=\; \mathcal{J} \,+\, \mathcal{J}_a \partial_a \,+\, \mathcal{J}_{ab} \,\partial_{ab} \,+\, \ldots \;.
\end{equation}
where $a...c$ denotes combinations of any number of indices and $\partial_{a...c} = \partial_a \cdots \partial_c$. 
Here repeated indices are summed over. Then the equations of motion for $\mathcal{J}$ read:
\begin{equation}
\label{eq:DofJ}
{\cal D}(\mathcal{J}_{a...c})\,\partial_{a...c} = \tau_d \,\partial_d  \Big( \mathcal{J}_{e...g} \,\partial_{e...g} \Big)\;,
\end{equation}
\eref{eq:DofJ} can be interpreted as a set of differential equation for the coefficients $\mathcal{J}_{a...c}$ since the differential basis is linearly independent.

To proceed further we will introduce perturbation theory and assume that $\tau$ is small. Since $\tau$ is related to the gravitational potentials this is usually a good approximation. For lensing it corresponds to the lensing being ``weak". We set the zero-th order $J^{(0)}$ to be unity. At the first order in $\tau$, we get
\begin{equation}
\Big[ {\cal{D}} (\mathcal{J}_a^{(1)}) \Big]\partial_a = \tau_a \partial_a.
\end{equation}
At the second order, we obtain two equations:  
\begin{equation}
\Big[ {\cal{D}} (\mathcal{J}_a^{(2)}) \Big] \partial_a 
= \Big[ \tau_d (\partial_d J_a^{(1)}) \Big] \partial_a,
\label{second1}
\end{equation}
and 
\begin{equation}
\Big[ {\cal{D}} (\mathcal{J}_{ab}^{(2)}) \big] \partial_a \partial_b 
= \Big[ \tau_a  J_b^{(1)} \Big] \partial_a \partial_b.
\label{second2}
\end{equation}
Note that the components of $\mathcal{J}$ with different number of indices do not evolve separately, but are coupled to each other. This is due to the derivative $\partial_d$ contained in the source $\tau$. It may either act on ${\mathcal J}_b$ or be part of the differential basis contributing to the component of ${\mathcal J}$ with one additional index, generating the two equations at the second order. 

To make this explicit in our notation we split $\,\partial_d = \partial_d^{\rightarrow} + \partial_d^{\uparrow}\,$, where $\partial_d^{\rightarrow} $ acts only on the coefficients of the differential basis and $ \partial_d^{\uparrow} $ acts only on the differential basis. For example, we define $\,\partial_d^{\rightarrow} (\mathcal{A}_{a...c}\,\partial_{a...c}) = 
\partial_d (\mathcal{A}_{a...c}) \,\partial_{a...c}\,\,$ and  $\,\,\partial_d^{\uparrow}(\mathcal{A}_{a...c}\,\partial_{a...c})
=  \mathcal{A}_{a...c} \,\partial_{a...cd}\,$.
Then to $n$-th order we get:
\begin{equation}
{\cal D}(\mathcal{J}^{(n)}_{a...c})\,\partial_{a...c} \:=\: \tau_d (\partial_d^{\rightarrow} + \partial_d^{\uparrow})\, \mathcal{J}^{(n-1)}_{e...g} \,\partial_{e...g}\;,
\end{equation}
which can be integrated in a straightforward way. We only need to be careful integrating the derivative operators, which may be time-dependent: $\partial_d^{\rightarrow}$ needs to be integrated over as it is a part of the differential equation for the coefficients (e.g. \eref{second1}), while $\partial_d^{\uparrow}$ is considered to be a part of the basis and is not part of the differential equation (e.g. \eref{second2}).
We obtain the formal solution: 
\begin{eqnarray}
\mathcal{J}^{(n)}_{a...c}(\eta_0) \,\partial_{a...c}&=& \int \limits_{\eta_{ini}}^{\eta_0} d\eta_1\, e^{-n^i \frac{\partial}{\partial x^i}(\eta_0-\eta_1)}\,\tau_d(\eta_1) \,\left[\;\partial_d^{\rightarrow}(\eta_1) + \partial_d^{\uparrow}(\eta_0)\,\right]\, \mathcal{J}^{(n-1)}_{e...g}(\eta_1) \,\partial_{e...g} \;,\\ \label{eq:JtoJminusone}
\mathcal{J}^{(n)}(\eta_0)&=& \int \limits_{\eta_{ini}}^{\eta_0} d\eta_1\, e^{-n^i \frac{\partial}{\partial x^i}(\eta_0-\eta_1)}\,\tau_d(\eta_1)\,\left[\;\partial_d^{\rightarrow}(\eta_1) + \partial_d^{\uparrow}(\eta_0)\,\right]\, \mathcal{J}^{(n-1)}(\eta_1) \;,
\end{eqnarray} 
where time arguments apply only to the coefficients and not the basis, which is always to be evaluated at $\eta_0$. We assumed initial conditions $\mathcal{J}(\eta_{ini}) = 1$, which means that $f$ and $\tilde{f}$ coincide at initial time, but then evolve differently. We are free to choose any initial condition for $\mathcal{J}$ as the removal of the non-linear propagation terms is independent of this choice. We will come back to the issue of initial conditions in the next section. 

We iterate \eref{eq:JtoJminusone} to relate $\mathcal{J}^{(n)}$ to $\mathcal{J}^{(0)}= 1$:
\begin{eqnarray} \nonumber
\mathcal{J}^{(n)}(\eta_0)&=& \int \limits_{\eta_{ini}}^{\eta_0} d\eta_1\, e^{-n^i \frac{\partial}{\partial x^i}(\eta_0-\eta_1)}\,\tau_{a_1}(\eta_1) \,\left[\;\partial_{a_1}^{\rightarrow}(\eta_1) + \partial_{a_1}^{\uparrow}(\eta_0)\,\right] \\ \nonumber
&&\times\int \limits_{\eta_{ini}}^{\eta_1} d\eta_2\, e^{-n^i \frac{\partial}{\partial x^i}(\eta_1-\eta_2)}\,\tau_{a_2}(\eta_2) \,\left[\;\partial_{a_2}^{\rightarrow}(\eta_2) + \partial_{a_2}^{\uparrow}(\eta_0)\,\right] \\ \nonumber
&& \hspace{3cm}\vdots
\\ \label{eq:Jfull}
&&\quad\times\int \limits_{\eta_{ini}}^{\eta_{n-1}} d\eta_n\, e^{-n^i \frac{\partial}{\partial x^i}(\eta_{n-1}-\eta_n)}\,\tau_{a_n}(\eta_n) \,\partial_{a_n}(\eta_0) \;.
\end{eqnarray}
This operator contains all propagation effects up to $n$-th order in $\tau$.
The structure simplifies in Fourier space as follows
\begin{eqnarray} \nonumber
\mathcal{J}^{(n)}(\eta_0,k_0)&=& \int \limits_{\eta_{ini}}^{\eta_0} d\eta_1\, \int \frac{d^3k_1 d^3k_1'}{(2\pi)^6}\,(2\pi)^3\, \delta(k_0 -k_1 -k_1') \,e^{- i n^i k_{0\,i}(\eta_0-\eta_1)}\\ \nonumber
&&\;\;\;\; \times\; \tau_{a_1}(\eta_1,k_1')\, \left[\;\partial_{a_1}^{\rightarrow}(\eta_1,k_1) + \partial_{a_1}^{\uparrow}(\eta_0)\;\right] \\ \nonumber
&& \hspace{3cm}\vdots
\\ \nonumber
&&\;\;\times \int \limits_{\eta_{ini}}^{\eta_{n-1}} d\eta_n\, \int \frac{d^3k_n d^3k_n'}{(2\pi)^6}\,(2\pi)^3\, \delta(k_{n-1} -k_n -k_n') \,e^{- i n^i k_{n-1 \, i}(\eta_{n-1}-\eta_n)} \\
&& \qquad\quad\; \times\; \tau_{a_n}(\eta_n,k_n') \, \partial_{a_n}(\eta_0)\,(2\pi)^3\,\delta(k_n)\;,
\end{eqnarray}
with $\frac{\partial}{\partial x^i}$ replaced by $i k$ in $\partial_{a}^{\rightarrow}(\eta,k)$. Note that we have not expanded the differential basis into Fourier space yet. 
The integrations can be simplified by disentangling the coupling of the time arguments in the integrations. We commute the factors $ e^{- i n^i k(\eta_a-\eta_b)}$ and obtain:
\begin{eqnarray}\nonumber
\mathcal{J}^{(n)}(\eta_0,k_0)&=& \int \limits_{\eta_{ini}}^{\eta_0} d\eta_1\, \int \frac{d^3k_1 d^3k_1'}{(2\pi)^6}\,(2\pi)^3\, \delta(k_0 -k_1 -k_1') \,e^{- i n^i k_{1\,i}'(\eta_0-\eta_1)}\\ \nonumber
&&\;\;\;\; \times\; \tau_{a_1}(\eta_1,k_1') \,\left[\;\partial_{a_1}^{\rightarrow}(\eta_1,k_1) + \partial_{a_1}^{\uparrow}(\eta_0) + ik_{1\,a_1} (\eta_0 - \eta_1)\;\right] \\ \nonumber
&&\;\times\int \limits_{\eta_{ini}}^{\eta_1} d\eta_2\, \int \frac{d^3k_2 d^3k_2'}{(2\pi)^6}\,(2\pi)^3\, \delta(k_1 -k_2 -k_2') \,e^{- i n^i k_{2\,i}'(\eta_0-\eta_2)} \\ \nonumber
&& \qquad\; \times\; \tau_{a_2}(\eta_2,k_2') \,\left[\;\partial_{a_2}^{\rightarrow}(\eta_2,k_2) + \partial_{a_2}^{\uparrow}(\eta_0)+ ik_{2\,a_2} (\eta_0 - \eta_2)\;\right]\\ \nonumber
&& \hspace{3cm}\vdots
\\ \nonumber
&& \quad\;\times\int \limits_{\eta_{ini}}^{\eta_{n-1}} d\eta_n\, \int \frac{d^3k_n d^3k_n'}{(2\pi)^6}\,(2\pi)^3\, \delta(k_{n-1} -k_n -k_n') \,e^{- i n^i k_{n\,i}'(\eta_{0}-\eta_n)} \\
\label{eq:J_fourier_space}
&&\qquad\qquad  \times\; \tau_{a_n}(\eta_n,k_n')\, \partial_{a_n}(\eta_0)\,(2\pi)^3\,\delta(k_n)\;,
\end{eqnarray}
with $\tau_a k_a = \tau_n^i k_i$ from the commutation of the exponentials with the derivative operators. 
If we consider a picture of the linear CMB at recombination, with all effects relocated to the last scattering surface, neglect time-delay and redshift terms and use a Newtonian approximation for the gravitational potentials, we find that the above formula is equivalent to the result obtained in Eq.~(62) of Ref.~\cite{su:2014a}.
In this simplified case that includes only lensing, our $\tau$ can be replaced by a derivative of the lensing potential, as we shall see in \eref{eq:lensing}.
The action of the derivative in Ref.~\cite{su:2014a} on the previous lenses generating the lens-lens coupling is equivalent to $\,\partial_{a_1}^{\rightarrow}(\eta_1,k_1) + ik_1 (\eta_0 - \eta_1)\,$ in our formula, while its action on the source is equivalent to our $\partial_{a_1}^{\uparrow}(\eta_0)$. Apart from notational differences, the two approaches are identical. Neglecting lens-lens couplings, all time integrations are independent and the remapping approach is rediscovered (for details see Ref.~\cite{su:2014a}). The advantage of our approach is that it is more general, and can account for time-delay and redshift terms in the same framework. In the following section we shall also see that we can add collisions in a natural way.

\section{Collisions} \label{Collisions} 
The procedure described above can be used to remove propagation effects from the Boltzmann equation also in the presence of a collision term $C(f)$. In this case, we obtain the following equation of motion for $\tilde{f}$:
\begin{equation}
D\tilde{f}  = \mathcal{J}^{-1}C(\mathcal{J}\tilde{f})\;.
\end{equation}
That is, the collision term for $\tilde{f}$ is evaluated using the fully lensed distribution function $\,\mathcal{J}\tilde{f} = f\,$, but has to be ``delensed'' by multiplication with $\mathcal{J}^{-1}$.
Using the line-of-sight integration and splitting the collision term as
$\,C(f) = -\dot{\kappa} f + \chi(f)\,$, where $\dot{\kappa}$ is 
the Compton scattering rate, we obtain:
\begin{equation}
\tilde{f}(\eta_0) = \int \limits_{\eta_{ini}}^{\eta_0} d\eta\, e^{-n^i\frac{\partial}{\partial x^i}(\eta_0-\eta)-\kappa} \mathcal{J}^{-1}(\eta) \,\chi(\mathcal{J}\tilde{f}(\eta))\;.
\end{equation}
The full distribution function today can be obtained by transformation with $\mathcal{J}(\eta_0)$:
\begin{equation}
f(\eta_0) = \mathcal{J}(\eta_0)\int \limits_{\eta_{ini}}^{\eta_0} d\eta\, e^{-n^i\frac{\partial}{\partial x^i}(\eta_0-\eta)-\kappa} \mathcal{J}^{-1}(\eta) \,\chi(f(\eta))\;.
\end{equation}
Applying $\mathcal{J}^{-1}(\eta)$ to the collision term cancels contributions from propagation effects prior to the source which is evaluated at the time $\eta$. This is necessary since $\mathcal{J}(\eta_0)$ is completely independent of the source and includes all space-time effects before $\eta_0$. 

It is possible to modify a second-order Boltzmann code such as \SONG to obtain $\mathcal{J}^{-1}$ directly by solving for $\mathcal{J}$ in cosmological perturbation theory at early times. This corresponds to adding an additional hierarchy to \SONG. It should be noted that in most cases $\mathcal{J}$ is needed one order below the target precision as it multiplies quantities of at least first order in the collision term \footnote{For \SONG,  which computes second-order perturbations, $\mathcal{J}$ is only needed at the linear level for lensing and time-delay. Only the redshift effects, which act on the background distribution function, have to be included at second order.}. As most time is spent computing the highest order, the performance of numerical codes is almost unaffected by this addition. Another importend numerical advantage is that this formalism eliminates the need to evaluate the unbound sum in $\ell$ appearing in the sources along the line-of-sight. After the transformation, the Boltzmann equation is source-free in the absence of collisions. Instead of evaluating the sum over $\ell$ at every numerical time step in the sources it is now sufficient to evaluate this sum only once, when the operator $\mathcal{J}$ is applied to $f$. Including collisions at recombination is unproblematic, as large multipole moments are still suppressed. For reionisation we refer to \cite{saito:2014a}; as $\mathcal{J}$ depends only on space-time and is independant of the distribution function, the appearance of unbound sums in $\ell$ can always be avoided.     

The important achievement of this formula is to create a split between perturbative collisional effects and non-perturbative line-of-sight effects. While a second-order code can compute all the physical effects in the early Universe ($ \mathcal{J}^{-1}(\eta) \,\chi(f(\eta))$) in cosmological perturbation theory, the late-time physics contained in $\mathcal{J}(\eta_0)$ can be computed in a different perturbative scheme using different tools. This formula not only clarifies the connection of lensing in the remapping approach to lensing at second order in the Boltzmann equations, it also specifies how to combine them using different methods suited for different epochs.

So far we assumed that the operator $\mathcal{J}$ is set to unity at the initial time. However our formalism does not depend on this choice and we can choose different initial conditions with each choice representing a different set of transformations that remove the propagation terms. We could set the initial conditions at the recombination $\eta_{rec}$ when the propagation effects become relevant. This choice leads to the minimal modification of the collision term as $\mathcal{J}^{-1}$ is very close to unity around recombination. However, if recombination is not instantaneous the choice of time is ambiguous and a clear separation of collisional and propagation effects is not achieved. Below we discuss in detail a natural way to solve this problem.

To treat the finite width of recombination, we set the initial conditions after the end of inflation $\eta_{ini} = 0$ and include a collision term in the equations of motion for $\mathcal{J}$ suppressing it during tight coupling:
\begin{equation}
[D,\mathcal{J}]+\tau \mathcal{J} = -\dot{\kappa}\,(\mathcal{J}-1)\;.
\end{equation}
The factor $\mathcal{J}-1$ ensures that at the background level $f$ and $\tilde {f}$ remain identical, as $\mathcal{J}^{(0)}=1$. With this collision term, $\mathcal{J}$ will start deviating from unity only after recombination. Since adding this collision term is not necessary, we are not forced to apply it at reionisation. The best strategy is to use only collisions at recombination ($\dot{\kappa}_{rec}$) to modify the equations of motion for $\mathcal{J}$. Adding a collision term does not spoil the removal of the non-linear propagation terms, but it does complicate the collision term for the distribution function $\tilde{f}$. We find:
\begin{eqnarray}
  \label{eq:f_eta0}
f(\eta_0) &=& \mathcal{J}(\eta_0)\int \limits_{\eta_{ini}}^{\eta_0} d\eta\, e^{-n^i\frac{\partial}{\partial x^i}(\eta_0-\eta)-\kappa} \mathcal{J}^{-1}(\eta)\,\left[\; \chi(f(\eta)) +\dot{\kappa}_{rec}(1-\mathcal{J}^{-1})f(\eta)\;\right]\\
&\approx& \mathcal{J}(\eta_0)\int \limits_{\eta_{ini}}^{\eta_0} d\eta\, e^{-n^i\frac{\partial}{\partial x^i}(\eta_0-\eta)-\kappa}  \,\chi(f(\eta)) \;,
\end{eqnarray}
where the second line is expanded to first order in the modified collision term using $\chi=\dot{\kappa}f^{(0)}+ \chi^{(1)}$ and $\mathcal{J}^{-1} = 1 - \mathcal{J}^{(1)}$, with $\mathcal{J}^{(0)} = 1$ at background level. This approximation is very accurate around recombination as at early times non-linear corrections are small. Note that it cannot be applied to reionisation, as the cancelation depends on $\mathcal{J}^{(1)}\dot{\kappa}_{\text{rec}}f^{(0)}$ cancelling with $\mathcal{J}^{(1)}\chi^{(0)}$, which by construction only works at recombination.

At the non-linear level using the full equation \eref{eq:f_eta0}, the term $\,\dot{\kappa}_{rec}(1-\mathcal{J}^{-1})f(\eta)$ takes into account the finite width of recombination. 
For our choice of initial conditions, the operator $\mathcal{J}^{-1}(\eta)$ singles out the effects of space-time on the distribution function around recombination, while $\mathcal{J}(\eta_0)$ describes the non-trivial structure of space-time and its effect on the distribution function after recombination \footnote{We would like to thank the authors of Ref. \cite{saito:2014a} for discussions on the physical interpretation of this transformation operator as a geodesic in inhomogeneous space-time.}.
This formalism allows to identify the residual terms of the propagation effects to any order in perturbation theory; these terms can be implemented in CMB codes like \SONG by modifying the collision term.
The calculation of $\mathcal{J}$ including the collision term is straightforward: only the factors $e^{-n^i\frac{\partial}{\partial x^i}(\eta_a-\eta_b)}$ have to be replaced by $e^{-n^i\frac{\partial}{\partial x^i}(\eta_a-\eta_b)-\kappa(\eta_a,\eta_b)}$ in \eref{eq:Jfull}.

In the remapping approach all effects have to be relocated to the surface of last scattering, which is inaccurate for the computation of reionisation and of the integrated Sachs-Wolfe (ISW) effect. Our method allows a natural inclusion of reionisation as there is no assumption on the time when the sources are active. In that case the factor $\mathcal{J}^{-1}(\eta)$ provides the needed corrections to the remapping approach.

An alternative approach is to set the initial condition today $\mathcal{J}(\eta_0)=1$ and solve $\mathcal{J}$ backwards in time. In that way the operator $\mathcal{J}$ multiplying the perturbations today is trivial.
The price to pay is a large modification of the collision term, which does contain the whole effect.
While we were working on this paper another group has developed a similar approach by using the line-of-sight integration along the full geodesics \cite{saito:2014a}. It turns out that for this choice of initial conditions the equations in the two approaches are identical. The operator ${\cal J}^{-1}$ in this case contains all information of the space-time between the time of photon emission and today, and takes care of the modification of the photon path due to space-time effects. As this approach is discussed in detail in their paper, we will now focus on the first option of fixing $\mathcal{J}$ at the time of recombination.

\section{Applications} \label{Applications}
In this chapter we calculate $\mathcal{J}$ to leading order in perturbation theory, assuming that $\tau$ is small. Note that we assume two approximations here. First we need to expand $\tau$ in perturbation theory, which does not assume that $\tau$ integrated along the line-of-sight is small, but that it is small at any point in space-time. Then, computing $\mathcal{J}$, we only consider a certain order of perturbation theory in $\tau$ itself which is related to the smallness along the line-of-sight. If the non-linear Newtonian potentials are used to compute $\tau$, the first order in $\tau$ will already provide very good results and only miss small corrections due to the lens-lens coupling. Even first order in cosmological perturbation theory, using the linear potential, will be sufficient to account for about 90\% of the lensing signal.

We now compute the leading order results in $\tau$ and in cosmological perturbation theory (or Newtonian theory). We use the longitudinal gauge to express the perturbed metric:
\begin{equation}
ds^2 = a(\eta)^2 \Big(- (1+2 A) \,d \eta^2 + (1 + 2 D) \,\delta_{ij}\, dx^i dx^j \Big).
\end{equation}
By comparing the Boltzmann equation in \eref{eq:boltzmann} with the definition of $\tau$ in \eref{eq:tau_parametrisation}, and by using the geodesic equation at first order, we find for the sources:
\begin{eqnarray}
\tau_{x}^i &=& n^i\:(A-D), \\
\tau_p &=& -p \,\frac{\partial}{\partial x^i}\,n^i A \:-\: p\,\dot{D}, \\
\tau_n^i &=& \sigma^{ij}\, \frac{\partial}{\partial x^j}\: (D-A)\;,
\end{eqnarray}
with the screen projector $\sigma^{ij} = \delta^{ij}-n^in^j$ and the potentials computed either linear in cosmological perturbation theory or in non-linear Newtonian theory. 

At leading order, $\mathcal{J}^{(1)}$, the different contributions do not mix, while at higher order effects such as lens-lens couplings and the lensing of redshift terms are important.
Here, we focus on the redshift and lensing terms and do not discuss the time-delay contributions, whose calculation follows in a straightforward way.
 
\subsection{Redshift terms}
\label{sec:redshift}
Using \eref{eq:J_fourier_space}, we find the following expression for the redshift terms:
\begin{eqnarray}
\mathcal{J}^{(1)}_p(\eta_0,k_0)&=&   \left( \int \limits_{0}^{\eta_0} d\eta_1 \,e^{- i n^i k_{0\,i}(\eta_0-\eta_1) - \kappa(\eta_0,\eta_1)}\,\tau_{p}(\eta_1,k_0) \right)\frac{\partial}{\partial p } \;.
\end{eqnarray}
This integration is identical to computing the ISW and Sachs-Wolfe (SW) effects in linear perturbation theory. After partial integration of the sources we obtain the ISW from the integration along the line-of-sight and the SW as the boundary term at recombination. We define $\theta_{\text{ISW}}$ as the CMB temperature perturbation induced by the SW and ISW effect:
\begin{equation}
\theta_{\text{ISW}} (\eta) 
= \int^{\eta}_0
d \eta_1 e^{- i n^i k_{0\,i}(\eta-\eta_1) - \kappa(\eta,\eta_1)} \left[ \:\dot{\kappa} A + (\dot{D} - \dot{A}) \,\right].
\end{equation}
Then 
\begin{equation}
\mathcal{J}_p(\eta_0) \;=\; 1\,-\,\theta_{\text{ISW}}\:p\,\frac{\partial}{\partial p }\;.
\end{equation}
The CMB perturbations including the redshift term can be obtained using \eref{eq:f_eta0}:
\begin{eqnarray}
f(\eta_0) &=& \mathcal{J}_p(\eta_0)\int \limits_{\eta_{ini}}^{\eta_0} d\eta\, e^{-n^i\frac{\partial}{\partial x^i}(\eta_0-\eta)-\kappa} \:\mathcal{J}_p^{-1}(\eta) \,\left[\:\chi(f(\eta)) +\dot{\kappa}_{rec}(1-\mathcal{J}_p^{-1})f(\eta)\,\right]\;.
\end{eqnarray}
Usually the presence of $\mathcal{J}^{-1}$ only adds a small correction since it is close to unity due to the tight-coupling suppression, but for the redshift term $\mathcal{J}$ receives a boundary contribution from the SW effect. However, it turns out that the corrections cancel exactly if the collision term is expanded to first order in perturbation theory, and we obtain:
\begin{eqnarray}
  \nonumber
f(\eta_0) &=& \mathcal{J}_p(\eta_0)\int \limits_{\eta_{ini}}^{\eta_0} d\eta\, e^{-n^i\frac{\partial}{\partial x^i}(\eta_0-\eta)-\kappa} \,\chi(f(\eta)) \\[0.3cm]
  \nonumber
&=& \mathcal{J}_p(\eta_0) \: f_\text{coll}(\eta_0) \;=\; (1-\theta_{\text{ISW}} \:p\,\frac{\partial}{\partial p })\: f_\text{coll} \\[0.3cm]
  \label{eq:redshift_f}
&\approx& f^{(1)}_\text{coll} \,-\, \theta_{\text{ISW}} \:p\,\frac{\partial}{\partial p }\: f^{(0)}\;,
\end{eqnarray}
with $f_\text{coll}$ the distribution function induced by only the collisional effects. To leading order in cosmological perturbation theory we rediscover the usual picture of the CMB, including collisional effects and the ISW plus SW effects. 


An alternative method to treat the numerically problematic part of the redshift terms which mixes metric and photon perturbations and involves unbound sums over $\ell$ was recently proposed in Ref.~\cite{huang:2014a} and later generalised to the polarised case in Ref.~\cite{fidler:2014a}.
The method employs the variable transformation $\,\tilde{\Delta} = \Delta -  \Delta\Delta/2\,$ at second order to remove the redshift terms involving the photon perturbations from the Boltzmann equation, with $\Delta$ the brightness
\begin{equation}
1 + \Delta \;=\; \frac{\int dp \,p^3 \,f(x,p)}{\int dp \,p^3 \,f^{(0)}(p)}\;.
\end{equation}
We use the method presented in this paper to derive a similar transformation. With $f = \mathcal{J} \tilde{f}$ and $\mathcal{J}_p \;=\; 1\,-\,\theta_{\text{ISW}}\:p\,\frac{\partial}{\partial p }$ we find the relation
\begin{equation}
\Delta = \tilde{\Delta} +\Delta_{\text{ISW}} + \Delta_{\text{ISW}}\tilde{\Delta}\,,
\end{equation} 
with $\Delta_{\text{ISW}}$ the brightness induced from the SW and ISW effect. 
In this transformation the quadratic terms $\Delta_{\text{ISW}}\tilde{\Delta}$ are responsible for removing the mixed redshift terms, while $\Delta_{\text{ISW}}$ removes the pure metric part. Note that here we have only considered the transformation $\mathcal{J}$ up to first order in $\tau_p$, which is sufficient for the mixed redshift terms as they multiply $\Delta$ which is itself at least of first-order in perturbations. 

To reproduce the results in Ref.~\cite{huang:2014a} one only needs to remove the numerically challenging mixed redshift terms and can therefore use the simpler transformation:
\begin{equation} \label{eq:trafoNew}
\tilde{\Delta} = \Delta - \Delta_{\text{ISW}}\Delta\,.
\end{equation} 
This transformation does not affect the pure metric part of the redshift terms, which is included as a source along the line-of-sight. The difference between the  transformation in \eref{eq:trafoNew} and the one in Ref.~\cite{huang:2014a} is in the modified collision term, which receives only minor corrections using the new transformation\footnote{The new transformation also requires adding the redshift-redshift correlation to the line-of-sight sources, which are numerically well behaved as they involve only the lowest multipoles. It is possible to avoid this by adding a further term to the transformation: $
\tilde{\Delta} = \Delta - \Delta_{\text{ISW}}\Delta + \frac{\Delta_{\text{ISW}}\Delta_{\text{ISW}}}{2}$.}.
The advantage of the new transformation is improving the numerical stability. The collision term is almost unchanged, while in the transformation of Refs.~\cite{huang:2014a, fidler:2014a} the collision term receives major corrections. This leads to a contribution from the surface of last scattering which is later cancelled by reverting the transformation at the final time $\eta_0$. With the new transformation there is no such cancelation leading to improved numerical performance. The method presented in this paper is preferable especially when treating polarisation, in which case the fact that the ISW is completely unpolarised greatly simplifies the transformation.


\subsection{Lensing terms}
For the lensing terms, \eref{eq:J_fourier_space} yields:
\begin{eqnarray}\nonumber
\mathcal{J}^{(1)}(\eta_0,k_0)&=&   \left( \int \limits_{0}^{\eta_0} d\eta_1\, e^{- i n^i k_{0\,i}(\eta_0-\eta_1) - \kappa(\eta_0,\eta_1)}\,\tau_{n}^i(\eta_1,k_0) \right)\left(\frac{\partial}{\partial n^i} + \eta_0 \frac{\partial}{\partial x^i}\right) \\
&&-\;\left( \int \limits_{0}^{\eta_0} d\eta_1\, e^{- i n^i k_0(\eta_0-\eta_1) - \kappa(\eta_0,\eta_1)}\;\eta_1\;\tau_{n}^i(\eta_1,k_{0\,i}) \right) \frac{\partial}{\partial x^i}\;.
\end{eqnarray}
The more complicated structure is due to the angular derivative operator $\bar{D}_i$ used for the lensing terms, defined in \eref{eq:tau_parametrisation}. Under the assumption that all CMB sources are located at the surface of last scattering, we can simplify the equation by enforcing $\frac{\partial}{\partial x^i}= \frac{1}{(\eta_{rec} - \eta_0)}\frac{\partial}{\partial n^i}$:
\begin{eqnarray} \label{eq:lensing}
\mathcal{J}^{(1)}(\eta_0,k_0) &=&  -\left( \int \limits_{0}^{\eta_0} d\eta_1\,\frac{1}{\eta_0-\eta_1} \sigma^{ij} \frac{\partial}{\partial n ^j} e^{- i n^i k_{0\,i}(\eta_0-\eta_1)- \kappa(\eta_0,\eta_1)} \:(D-A) \right)\frac{\partial}{\partial n^i} \\ \nonumber
&&+\left( \int \limits_{0}^{\eta_0} d\eta_1\, \frac{1}{\eta_{0}-\eta_{rec}} \sigma^{ij} \frac{\partial}{\partial n ^j}  e^{- i n^i k_{0\,i}(\eta_0-\eta_1)- \kappa(\eta_0,\eta_1)} \:(D-A)\right) \frac{\partial}{\partial n^i} \\ \nonumber
&= &\frac{\partial}{\partial n^{\perp}_i} \left( \int \limits_{0}^{\eta_0} d\eta_1\, \frac{\eta_{rec}-\eta_1}{(\eta_{0}-\eta_{rec})(\eta_{0}-\eta_1)}    e^{- i n^i k_{0\,i}(\eta_0-\eta_1)- \kappa(\eta_0,\eta_1) } \:(D-A)\right) \frac{\partial}{\partial n^i_{\perp}}\;,
\end{eqnarray}
with $\frac{\partial}{\partial n_{\perp} ^i} = \sigma^{ij}\frac{\partial}{\partial n^j}$. This approximation is based on assuming that $\mathcal{J}$ acts on a function with $\vec{n}$-dependence of $\text{exp}(n^i \frac{\partial}{\partial x^i}(\eta_{\text{rec}}-\eta_0))$, resulting from a monopole source at recombination. Corrections to this approximation arise from the width of recombination and sources that are not localised at the surface of last scattering, e.g. the ISW effect. However, as these are usually late time effects they almost unaffected by lensing. In addition there is a small correction from the weak angular dependance of the sources at recombination. 
By going back to real space, the time integration yields the projected lensing potential along the line-of-sight:
\begin{equation}
\psi(n^i)
\;=\;  \int \frac{d^3k}{(2 \pi)^3} \int \limits_{0}^{\eta_0} d\eta_1\, \frac{\eta_{rec}-\eta_1}{(\eta_{0}-\eta_{rec})(\eta_{0}-\eta_1)}   \: e^{- i n^i k_i(\eta_0-\eta_1)- \kappa(\eta_0,\eta_1)}\, \Big(D(\eta_1,k)-A(\eta_1, k) \Big), 
\end{equation}
and we obtain:
\begin{eqnarray}
  \label{eq:lensing_f}
  \nonumber
f(\eta_0) &=& \mathcal{J}(\eta_0)\int \limits_{\eta_{ini}}^{\eta_0} d\eta\, e^{-n^i\frac{\partial}{\partial x^i}(\eta_0-\eta)-\kappa} \,\chi(f(\eta)) \\[0.3cm]
  \nonumber
&=& \mathcal{J}(\eta_0)\: f_\text{coll}(\eta_0) \\[0.3cm]
&\approx& f_\text{coll}  \,+\,\left(\frac{\partial}{\partial n_i^{\perp}} \psi(n^i)\right)\left(\frac{\partial}{\partial n^i_{\perp}} f_\text{coll}\right )\;.
\end{eqnarray}

As there are no correlations between the redshift and lensing terms at leading order in $\tau$, it is straightforward to combine both using \eref{eq:redshift_f} and \eref{eq:lensing_f}:
\begin{eqnarray}
  \label{eq:lensing_redshift_f}
f(\eta_0) &=& f^{(0)} \;+\; f^{(1)}_\text{coll}  \;-\; \theta_{\text{ISW}}\:p\,\frac{\partial}{\partial p} \,(f^{(0)} + f^{(1)}_\text{coll}) \:+\: \left(\frac{\partial}{\partial n_i^{\perp}} \psi^{(1)}(n^i)\right)\left(\frac{\partial}{\partial n^i_{\perp}} f^{(1)}_\text{coll}\right )\;.
\end{eqnarray} 
The second term describes the usual collision sources at linear order, while the third term contains the ISW and SW effect and the redshifting of the first-order collision part. The last term describes the lensing of the collisional CMB, without the lensing of the ISW and SW, which is included at higher order in $\tau$. Computing these higher-order contributions requires numerically challenging integrations along the line-of-sight; for details we refer to \cite{saito:2014a}. Note that the transport operator formalism also provides the residual terms of lensing and the redshift term at recombination that can be added using a Boltzmann code. Computing the time-delay transformation works in exactly the same way, without extra complications due to boundary terms or modified derivative operators.
This result is leading order in $\tau$ and leading oder in cosmological perturbation theory. Considering that lensing is a nonlinear effect, contributions to first oder in $\tau$ and first order in cosmological perturbations are large compared to other second order contributions. However, for a full analysis to second order in perturbation theory additional contributions from the non-linear collision term and the second-order redshift contributions in $\tau$, representing the second-order counterpart of the ISW, have to be considered. 
 
In this framework, we can compute collisional effects up to second order in cosmological perturbation theory using a second-order Boltzmann code such as \SONG. The hierarchy for $\mathcal{J}$ can be added and used to evaluate the residuals due to the finite width of recombination. Then, the lensing, time-delay and redshift effects can be added to a given order in $\tau$ using the non-linear Newtonian potentials.
For the first time, we have a consistent framework to treat all the effects contained in the Boltzmann equation, which also answers the question of how to combine the standard lensing calculations with existing Boltzmann codes at any order in perturbation theory.

As an application of the transport operator formalism, in Appendix A we compute the correlation between the temperature anisotropies caused by lensing and the gravitational redshift.

\section{Polarisation} \label{Polarisation}
In this section we briefly show how to extend our analysis to polarisation. To do so, we need to compute the transport of the polarisation tensor $\,f_{ab}\,$ in space-time, where $ab$ are helicity indices \cite{pitrou:2009b,beneke:2010a}, instead of just the scalar distribution function $f$. The transport terms for $f_{ab}$ are identical to the unpolarised analysis, only the angular derivatives are replaced by spin-raising and spin-lowering derivatives. To absorb the propagation effects we define a transformation operator $\mathcal{J}_{ab}$, but it turns out that only the diagonal components of $\mathcal{J}$ are relevant. The source $\tau$ is diagonal in the helicity indices and we can choose initial conditions $\mathcal{J}_{ab} = \text{diag}(\mathcal{J})_{ab}$. For the diagonal elements we recover the standard angular derivatives, effectively reducing the formula to the scalar $\mathcal{J}$ discussed in the previous sections. The operator $\mathcal{J}$ removing the propagation effects depends only on the space-time and is independent of the distribution function it is acting on.

The angular derivative $\frac{\partial}{\partial n^i_\perp}$ has to be replaced with a covariant derivative when acting on spin two quantities. Following Ref.~\cite{beneke:2010a} we substitute 
\begin{equation}
\frac{\partial}{\partial n^i_{\perp}} \;\rightarrow\; \frac{1}{\sqrt{2}}\left(\epsilon_-^i \eth_s + \epsilon_+^i\bar{\eth}_s \right)\;,
\end{equation}
where $\eth$ and $\bar{\eth}$ are the spin-raising and spin-lowering derivatives, $s$ the spin of the quantity the derivative is acting on (either $0$ or $\pm 2$) and $\epsilon_{\pm}$ is the helicity basis on the sphere.
Next we transform from $ab$ to the usual polarisation basis $X=I,E,B$. In multipole space we obtain:
\begin{eqnarray}
& f_{ab} &= \mathcal{J} \tilde{f}_{ab}  \nonumber \\
\Rightarrow & f_{ab,lm} &= i^{l-l'-l''} \sqrt{\frac{4\pi(2l+1)}{(2l'+1)(2l''+1)}}\int d\Omega \:Y_{lm}^{s*}(\vec{n})\, Y_{l''m''}(\vec{n})\, \mathcal{J}_{l''m''}\, Y_{l'm'}^{s}(\vec{n}) \,\tilde{f}_{ab,l'm'} \;.
\end{eqnarray}
We will now assume that $\mathcal{J}_{l'm'}$ does not contain angular derivatives acting on $Y_{l''m''}^s(\vec{n})$.
Then, the angular integration simply reduces to a Gaunt integral:
\begin{equation}
f_{ab,lm} \;=\; i^{l-l'-l''} \left(\begin{array}{cc|c}l'&l''&l\\m'&m''&m\end{array}\right)\left(\begin{array}{cc|c}l'&l''&l\\-s&0&-s\end{array}\right)\mathcal{J}_{l''m''}\tilde{f}_{ab,l'm'}\;.
\end{equation}
Here the appearance of the spin $s$ in the Clebsch-Gordan coefficients generates a mixing of E and B polarisation. Following Ref.~\cite{beneke:2010a}, in the basis $X=I,E,B$ we find:
\begin{equation}
f_{X,lm} \;=\; i^{l-l'-l''} H^*_{XY}(l-l'-l'')\left(\begin{array}{cc|c}l'&l''&l\\m'&m''&m\end{array}\right)\left(\begin{array}{cc|c}l'&l''&l\\F_X&0&F_X\end{array}\right)\mathcal{J}_{l''m''}\tilde{f}_{Y,l'm'}\;,
\end{equation}
where $H_{XY}$ is a simple matrix that mixes the E and B-modes of polarisation.

In presence of lensing, the assumption of no angular derivatives does not hold, and one has to resort to a full calculation. The angular integration is then more lengthy, as one needs to resort to spin-raising and spin-lowering derivatives; refer to Ref.~\cite{beneke:2010a} for details. However, the structure of the solution is unchanged. In particular, the mechanism of converting E into B polarisation stays the same. The complexity is reduced when considering a flat sky approach; the relevant calculations in this case are summarised in Ref.~\cite{lewis:2006a}.


\section{Conclusions} \label{Conclusions}

In this paper we introduced the transport operator formalism, a new framework to compute the non-linear CMB anisotropies.
The formalism is fully perturbative, it does not assume that all sources are located at the surface of last scattering, it includes redshifting, lensing and time-delay effects and it naturally incorporates polarisation.
As such, it generalises previous methods like the remapping approach for lensing \cite{hu:2000b, lewis:2006a} and those in Refs.~\cite{huang:2014a, su:2014a}.

The salient point of the transport operator formalism is the possibility to clearly separate the linear physics of the early Universe, mainly collisions from the time of recombination, from the non-linear propagation effects of the late Universe, such as lensing, redshifting and time-delay effects.
This separation is achieved via two perturbative expansions in the Boltzmann equation. The collisional physics is treated in the standard framework of cosmological perturbation theory, using for instance a second-order Boltzmann code such as \SONG.
The late-time physics instead is described in terms of a distinct perturbative expansion of the transport operator in \eref{eq:tau_definition} which, at leading order and when the sources are located at the surface of last scattering, naturally leads to the well known remapping approach for lensing, as can be seen in \eref{eq:lensing_redshift_f}.

In addition, our formalism provides the corrective terms arising from the finite width of recombination and for sources that are spread out in time, like reionisation and the integrated Sachs-Wolfe effect, at any perturbative order.
At next-to-leading order, the transport operator formalism correctly describes effects like lens-lens coupling and the lensing of the ISW and SW effects.
We plan to implement and quantify these effects in a future work by updating the second-order Boltzmann code \SONG.

\acknowledgments
We thank R. Saito, A. Naruko, T. Hiramatsu and M. Sasaki for useful discussions, especially on the connection between $\mathcal{J}$ and the line-of-sight integration along a full geodesic \cite{saito:2014a}, which helped us clarifying the physical interpretation of the transformation operator $\mathcal{J}$.  
C.~Fidler and K.~Koyama are supported by the UK Science and Technology Facilities Council grants number ST/K00090/1 and ST/L005573/1.
GWP acknowledges support by the UK STFC grant ST/I000976/1.
The research leading to these results has received funding from the European Research Council under the European Union's Seventh Framework Programme (FP/2007-2013) / ERC Grant Agreement No. [616170].

\appendix

\section{Redshift-Lensing Correlation} \label{Correlation}
\begin{figure}[t]
  \centering
    \includegraphics[width=0.8\textwidth,natwidth=550,natheight=400]{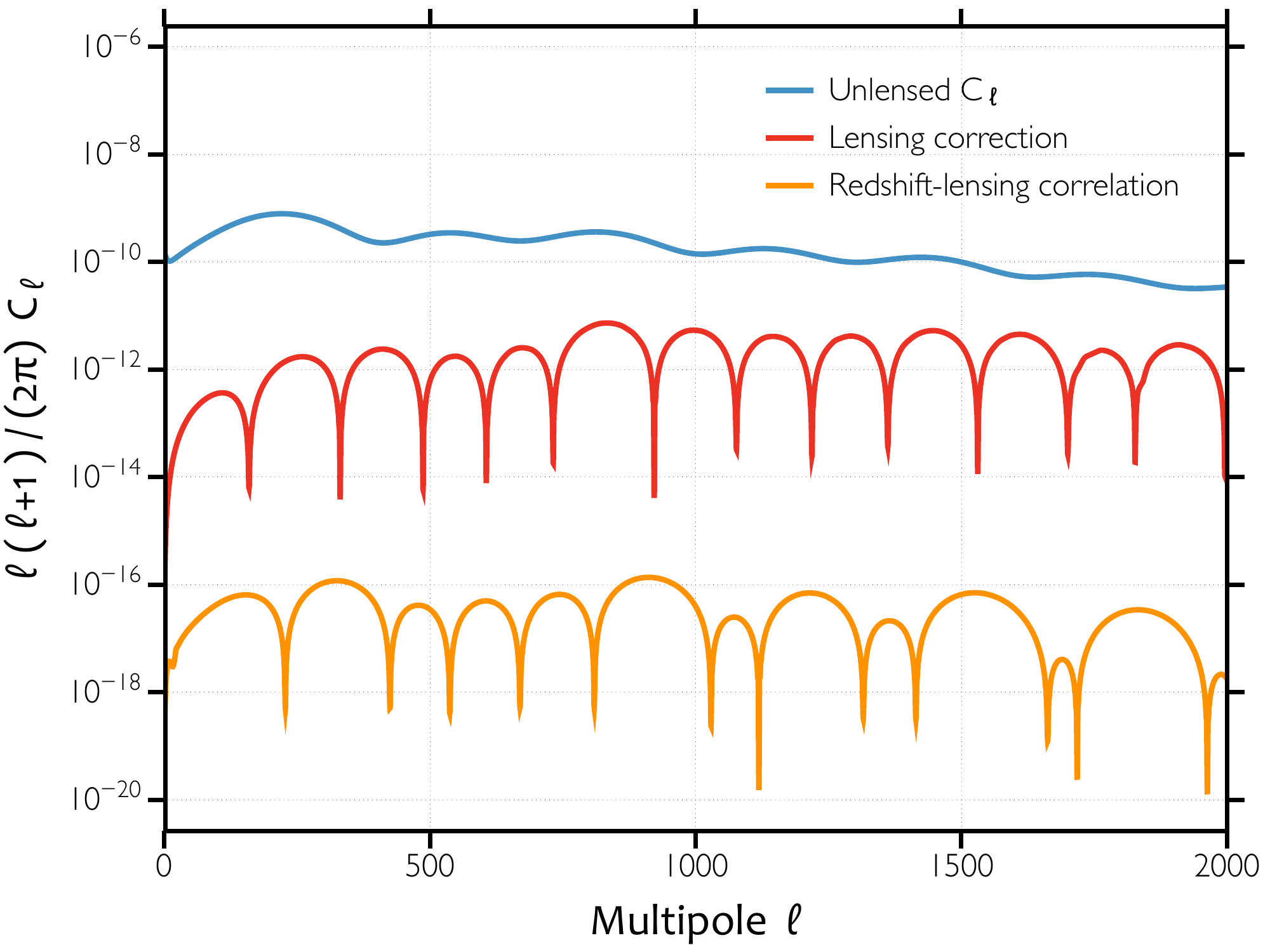}
  \caption{
From top to bottom, the linear intensity power spectrum compared to its correction due to lensing and the redshift-lensing correlation. On all examined scales the redshift-lensing correlation is a factor of $10^4$ smaller than the lensing correction. 
}
  \label{fig:lensing-redshift-correlation}
\end{figure}

As an application of the transport operator formalism we compute the contributions of the redshift terms to the power spectrum beyond the ISW and SW effects. Assuming a vanishing primordial non-Gaussianity, the next leading contribution is the correlation of the redshift terms with lensing. For simplicity we will perform all calculations using the flat-sky approach.
For lensing, \eref{eq:lensing_f} can be used in the flat-sky limit to find the standard result \cite{lewis:2006a}:
\begin{equation}
\theta_{\text{lensing}} \;=\; -\int \frac{dl'^2}{2\pi}\,l'(l-l')\:\psi_{|l-l'|}\:\theta_l\;.
\end{equation}
In \sref{sec:redshift}, we have found for the redshift terms
$\,\mathcal{J}_{l} = -\theta^{\text{ISW}}_{l}\,p\,\frac{\partial}{\partial p}\,$; using \eref{eq:redshift_f} in the flat-sky limit, it follows that the non-linear photon perturbations induced from the redshift terms are given by
\begin{equation}
\theta_{\text{redshift}}\;=\;4\int \frac{dl'^2}{2\pi}\;\theta^{\text{ISW}}_{|l-l'|}\;\theta_l\;.
\end{equation}
The lensing potential is independent from recombination effects as it only correlates with the ISW. Thus, the correction to the angular power spectrum due to redshift-lensing correlation is given by
\begin{equation}
C_l^{\text{lensing-redshift}} \;=\; -4 \int \frac{dl'^2}{(2\pi)^2}\;l'(l-l')\;C^{\psi \text{ISW}}_{|l-l'|}\;C^\theta_l\;.
\end{equation}
We compute this correction with the linear solver CLASS \cite{lesgourgues:2011a, blas:2011a} and find it to be a factor $10^{4}$ smaller than the lensing correction to the power spectrum, as can be seen by \fref{fig:lensing-redshift-correlation}, and comparable to other corrections to lensing such as lens-lens coupling \cite{su:2014a}.
Its effect may be larger for the bispectrum, as lensing is already limited to the correlation of the lensing potential with the ISW effect. The effect on the bispectrum has been previously computed \cite{pettinari:2013a, pettinari:2014b} using the $\Delta\Delta$-transformation introduced in \sref{sec:redshift}, and it was found that the redshift terms are indeed relevant for the bispectrum analysis.

\end{document}